\documentclass[11pt,twoside]{article}
\usepackage{asp2010}

\resetcounters

\begin{document}

\title{SCUBA-2 Data Processing}
\author{Tim~Jenness$^1$, David~Berry$^1$, Ed~Chapin$^2$,
  Frossie~Economou$^1$, Andy~Gibb$^2$, and Douglas~Scott$^2$
\affil{$^1$Joint Astronomy Centre, 660 N.\ A`oh\={o}k\={u} Place, HI, 96720, USA}
\affil{$^2$Department of Physics \& Astronomy, University of British Columbia, 6224 Agricultural Road, Vancouver, BC V6T 1Z1, Canada}}

\begin{abstract}
  SCUBA-2 is the largest submillimetre array camera in the world and
  was commissioned on the James Clerk Maxwell Telescope (JCMT) with
  two arrays towards the end of 2009. A period of shared-risks
  observing was then completed and the full planned complement of 8
  arrays, 4 at 850\,\micron\ and 4 at 450\,\micron, are now installed
  and ready to be commissioned. SCUBA-2 has 10,240 bolometers,
  corresponding to a data rate of 8\,MB/s when sampled at the nominal
  rate of 200\,Hz. The pipeline produces useful maps in
  near real time at the telescope and often publication quality maps in
  the JCMT Science Archive (JSA) hosted at the Canadian Astronomy
  Data Centre (CADC).
\end{abstract}

\section{SMURF Iterative Map-Maker}

\begin{figure}
\begin{center}
\includegraphics[angle=-90,width=0.9\textwidth]{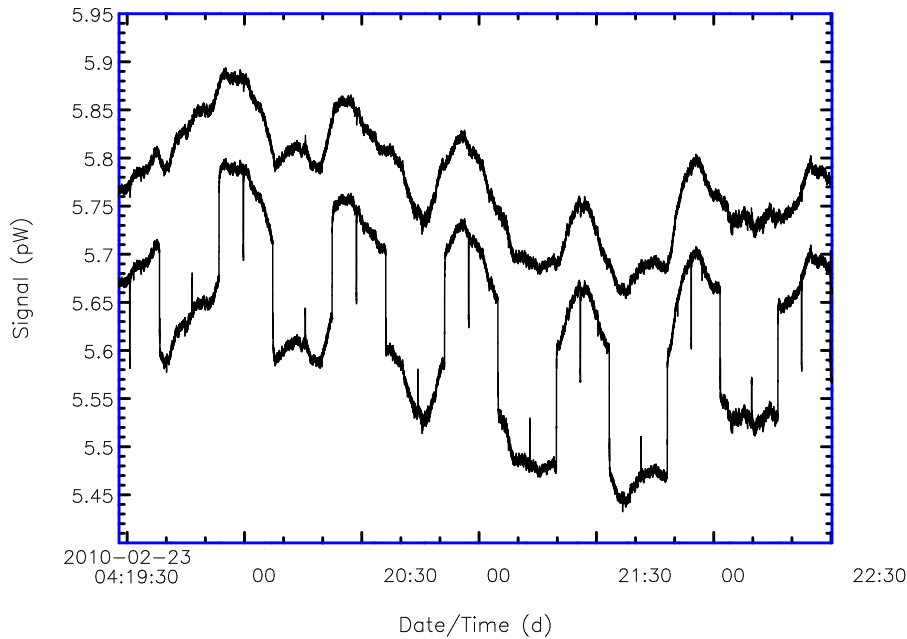}
\caption{The lower curve shows some flatfielded data containing many
  steps along with some spikes. The upper curve is the same data with
  the steps and spikes removed (including an offset of 0.15~pW to make
  this easier to see).}
\label{fig:steps}
\end{center}
\end{figure}

The Sub-Millimetre Common-User Bolometer Array 2 (SCUBA-2)
(\cite{2010SPIE.7741E..43C,2006SPIE.6275E..45H}) is a direct-detection
bolometer array and so measures the temperature variation of the sky
as well as the astronomical signal. Data are taken by scanning the
telescope over the source using a pattern designed such that the time
taken to return to the same place on the sky is not a fixed
interval. The telescope software has been modified to use a number of
different patterns called ``pong'', ``lissajous'' and ``daisy'' that
have this property (\cite{2010SPIE.7740E..66K}).  This allows the
map-maker to separate time-varying signals from those that are fixed
in a particular location on the sky. We have developed the SMURF
software package (Sub-Millimetre User Reduction Facility;
\cite{SMURF}) to process SCUBA-2 data. The SMURF map-maker works by
iteratively fitting a collection of models to the time stream in turn,
and subtracting them, leaving the astronomical signal. All model
components are refined at each iteration, resulting in improved
astronomical image estimates. However, before the iterations may
begin, we must repair several problems with the bolometer time series.
For example, the SQUID readout electronics can introduce steps which
need to be fixed. We have developed a robust algorithm for correcting
these problems and an example is shown in Fig.\ \ref{fig:steps}.

\subsection{Data Models}

After glitch repair, the time series data in Fig.\ \ref{fig:steps}
still have a periodic structure that is dominated by a 25 second
oscillation in the fridge along with low-frequency variations due to
changes in the sky power and instrument drifts. These signals are
common to all the bolometers and so can be removed, albeit with a loss
of sensitivity to signals larger than the array footprint. We reject
any bolometers that do not exhibit this strong common-mode signal. In
addition, the relative amplitudes of the common-mode in different
bolometers may be used to refine the flatfield.

Other models include a Fourier filter to remove low and high
frequencies from the data based on the scan speed of the telescope and
the wavelength, correction for atmospheric extinction, and an
alternative to the Fourier filter that removes a median value from a
rolling box.

\begin{figure}[t]
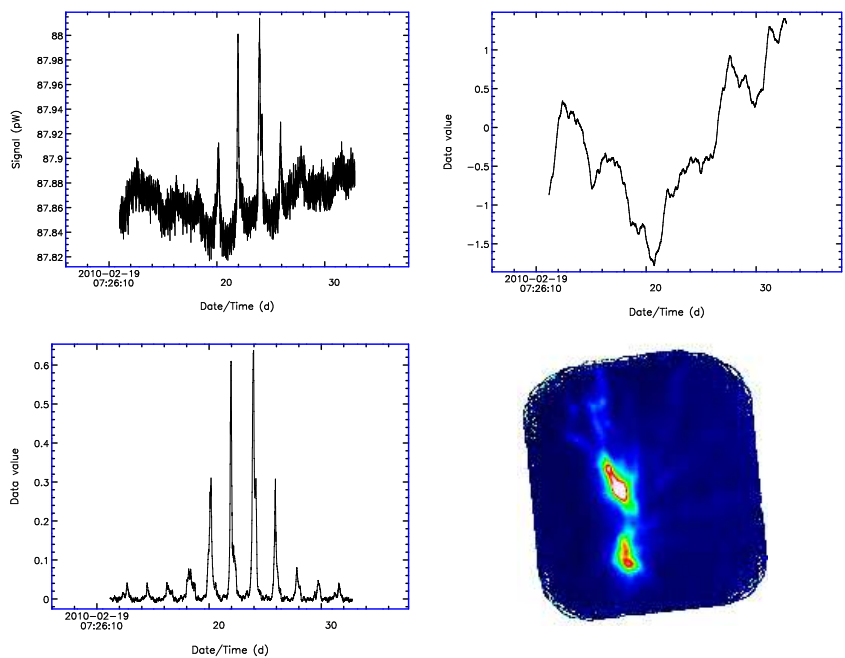

\begin{center}
\begin{tabular}{cc}
\includegraphics[angle=-90,width=0.4\textwidth]{f2a.eps} &
\includegraphics[angle=-90,width=0.4\textwidth]{f2b.eps} \\
\includegraphics[angle=-90,width=0.4\textwidth,clip]{f2c.eps}&
\includegraphics[angle=-90,width=0.25\textwidth,clip]{f2d.eps}
\end{tabular}

\caption{Top left shows some raw flatfielded time-series data from a single
  bolometer for a 20 second observation of a source in Orion. Top right is the
  common-mode signal and the bottom left is the astronomical signal
  determined from the time series after removing all the models. The
  final map is shown bottom right and covers an area of about
  4 arcmin $\times$ 4 arcmin.}
\label{fig:models}
\end{center}
\end{figure}

\begin{figure}[t]
\begin{center}
\includegraphics[width=0.75\textwidth]{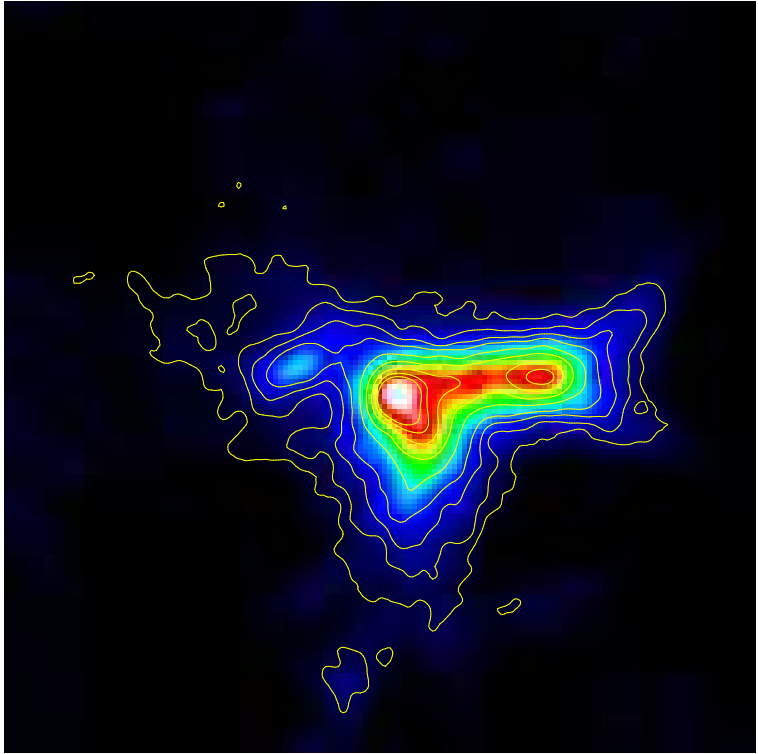}
\caption{Commissioning image of IRAS\,18592+0108 at 450\,\micron\ taken on
  2009 October 25th with contours from an 850\,\micron\ SCUBA
  scan map commissioning observation from 1997 August 19th (see also \cite{1996ESASP.388..135G}).}
\label{fig:w48}
\end{center}
\end{figure}

At the end of each iteration, once the low-frequency noise and
astronomical signal components have been estimated and removed, the
residual signal is considerably flatter making it easy to identify
smaller spikes. We have also implemented a map-based despiker that
identifies outliers in the data that land in each map pixel. This
procedure is repeated until the RMS of the residuals do not change
appreciably. Fig.\ \ref{fig:models} shows an example of some models
for a source in the Orion Molecular Cloud and Fig.\ \ref{fig:w48}
shows a commissioning observation compared with some commissioning
data from the first SCUBA taken in 1997.

\section{Pipeline and the JCMT Science Archive}

At the telescope and at the JSA (hosted at CADC) we use the ORAC-DR
data reduction pipeline (\cite{2008AN....329..295C}) to run the SMURF
map-maker and to perform mosaicking, pointing corrections and data
analysis. PiCARD (\cite{2008ASPC..394..565J}) is used for off-line
data analysis.

At the summit there are two pipelines, one for each wavelength, for
Quick Look, and two pipelines for science processing all running on
dedicated machines. The Quick Look pipeline
(\cite{2005ASPC..347..585G}) processes single observations as quickly
as possible to provide instant feedback to the observer and also to
reduce pointing and focus observations. The science pipelines have a
little more time to process the data and can demonstrate observation
progress by waiting for more data to accumulate and mosaicking
multiple observations.

In the JSA time is not an issue, so the pipeline can run on the CADC
grid processing cloud (\cite{O11_2_ADASSXX}) using more complex
models and many more iterations.  It is also possible to run the
pipeline in the new CANFAR cloud computing infrastructure
(\cite{P077_ADASSXX}).

\bibliography{P056}

\begin{thebibliography}{}
\expandafter\ifx\csname natexlab\endcsname\relax\def\natexlab#1{#1}\fi
\expandafter\ifx\csname url\endcsname\relax
  \def\url#1{\texttt{#1}}\fi
\expandafter\ifx\csname urlprefix\endcsname\relax\def\urlprefix{URL }\fi
\providecommand{\eprint}[2][]{\url{#2}}

\bibitem[{{Cavanagh} et~al.(2008){Cavanagh}, {Jenness}, {Economou}, \&
  {Currie}}]{2008AN....329..295C}
{Cavanagh}, B., {Jenness}, T., {Economou}, F., \& {Currie}, M.~J. 2008,
  Astronomische Nachrichten, 329, 295

\bibitem[{Chapin et~al.(2010)Chapin, Gibb, Jenness, Berry, \& Scott}]{SMURF}
Chapin, E., Gibb, A.~G., Jenness, T., Berry, D.~S., \& Scott, D. 2010, Starlink
  User Note 258, Joint Astronomy Centre

\bibitem[{{Craig} et~al.(2010)}]{2010SPIE.7741E..43C}
{Craig}, S.~C., et~al. 2010, in Society of Photo-Optical Instrumentation
  Engineers (SPIE) Conference Series, vol. 7741

\bibitem[{Economou et~al.(2011)}]{O11_2_ADASSXX}
Economou, F., et~al. 2011, in ADASS XX, edited by I.~N. Evans, A.~Accomazzi,
  D.~J. Mink, \& A.~H. Rots (San Francisco: ASP), vol. TBD of ASP Conf. Ser.,
  TBD

\bibitem[{Gaudet et~al.(2011)}]{P077_ADASSXX}
Gaudet, S., et~al. 2011, in ADASS XX, edited by I.~N. Evans, A.~Accomazzi,
  D.~J. Mink, \& A.~H. Rots (San Francisco: ASP), vol. TBD of ASP Conf. Ser.,
  TBD

\bibitem[{{Gear} et~al.(1996){Gear}, {Holland}, {Cunningham}, \&
  {Lightfoot}}]{1996ESASP.388..135G}
{Gear}, W.~K., {Holland}, W.~S., {Cunningham}, C.~R., \& {Lightfoot}, J.~F.
  1996, in Submillimetre and Far-Infrared Space Instrumentation, edited by
  {E.~J.~Rolfe \& G.~Pilbratt}, vol. 388 of ESA Special Publication, 135

\bibitem[{{Gibb} et~al.(2005){Gibb}, {Scott}, {Jenness}, {Economou}, {Kelly},
  \& {Holland}}]{2005ASPC..347..585G}
{Gibb}, A.~G., {Scott}, D., {Jenness}, T., {Economou}, F., {Kelly}, B.~D., \&
  {Holland}, W.~S. 2005, in Astronomical Data Analysis Software and Systems
  XIV, edited by {P.~Shopbell, M.~Britton, \& R.~Ebert}, vol. 347 of
  Astronomical Society of the Pacific Conference Series, 585

\bibitem[{{Holland} et~al.(2006)}]{2006SPIE.6275E..45H}
{Holland}, W., et~al. 2006, in Society of Photo-Optical Instrumentation
  Engineers (SPIE) Conference Series, vol. 6275

\bibitem[{{Jenness} et~al.(2008){Jenness}, {Cavanagh}, {Economou}, \&
  {Berry}}]{2008ASPC..394..565J}
{Jenness}, T., {Cavanagh}, B., {Economou}, F., \& {Berry}, D.~S. 2008, in
  Astronomical Data Analysis Software and Systems XVII, edited by
  {R.~W.~Argyle, P.~S.~Bunclark, \& J.~R.~Lewis}, vol. 394 of Astronomical
  Society of the Pacific Conference Series, 565

\bibitem[{{Kackley} et~al.(2010){Kackley}, {Scott}, {Chapin}, \&
  {Friberg}}]{2010SPIE.7740E..66K}
{Kackley}, R., {Scott}, D., {Chapin}, E., \& {Friberg}, P. 2010, in Society of
  Photo-Optical Instrumentation Engineers (SPIE) Conference Series, vol. 7740

\end{thebibliography}

\end{document}